\documentclass{article}
\usepackage{spconf,amsmath,graphicx, xcolor}
\usepackage{amsfonts, bm, bbm, mathtools, multirow, makecell, caption}
\usepackage{algorithm,algcompatible, setspace}
\title{Expressive Acoustic Guitar Sound Synthesis with an Instrument-Specific Input Representation and Diffusion Outpainting}
\name{Hounsu Kim, Soonbeom Choi, Juhan Nam
\thanks{This work was supported by the National Research Foundation of Korea (NRF) grant funded by the Korea government (MSIT) (No. RS-2023-00222383).}
}
\address{Graduate School of Culture Technology, KAIST, Daejeon, Republic of Korea}
\begin{document}
\ninept
\maketitle
\begin{abstract}

Synthesizing performing guitar sound is a highly challenging task due to the polyphony and high variability in expression. Recently, deep generative models have shown promising results in synthesizing expressive polyphonic instrument sounds from music scores, often using a generic MIDI input. In this work, we propose an expressive acoustic guitar sound synthesis model with a customized input representation to the instrument, which we call \textit{guitarroll}. We implement the proposed approach using diffusion-based outpainting which can generate audio with long-term consistency. To overcome the lack of MIDI/audio-paired datasets, we used not only an existing guitar dataset but also collected data from a high quality sample-based guitar synthesizer. Through quantitative and qualitative evaluations, we show that our proposed model has higher audio quality than the baseline model and generates more realistic timbre sounds than the previous leading work.

\end{abstract}
\section{Introduction}
\label{sec:introduction}

MIDI-to-audio synthesis is an essential process in producing music. 
Current off-the-shelf methods rely heavily on sample-based synthesis using prerecorded, high-quality audio.
However, sample-based synthesis is generally not flexible enough to handle the wide acoustic variability of expressive instruments. 
One such instrument is the acoustic guitar, where even a simple strum across six strings exhibits subtle variations in microtiming between note onsets and pick noise with each repetition. While recording all possible samples \cite{schwarz2000} or physically modeling them \cite{laurson2001realisticguitarsynth} is feasible, these approaches are limited in their ability to capture contextual variations and real-world performance nuances.

Recent advancements in deep generative models have opened up the possibility of faithfully modeling expressive sounds using large-scale audio data. Early work utilized likelihood-based loss functions \cite{manzelli2018, wang2019performancenet, kimjongwook2019, dong2022deepperformer}, 
but they encountered smoothing issues because they do not consider the variance of the data distribution \cite{mathieu2016mse}. On the other hand, the adversarial loss directly considers the distance between the distribution of the train set and predicted data by introducing a lower bound, which is expressed as a feasible expectation formation \cite{nowozin2016fgan}. However, existing neural audio synthesis works based on the adversarial loss \cite{engel2018gansynth, narita2023ganstrument} focus on synthesizing single MIDI notes utilizing the NSynth dataset \cite{engel2017}.

Recently, MIDI-DDSP \cite{wu2022mididdsp} has successfully disentangled audio into pitch, loudness and timbre features by integrating a traditional signal processing model into a differential neural architecture. This enables high-fidelity audio generation with good controllability when a MIDI sequence is given, but their methods only apply to monophonic cases. Deep performer \cite{dong2022deepperformer} generates polyphonic audio with expressive timing using a transformer-based model. Their work focuses on piano and violin instruments and receives musical score rather than MIDI as the model input. Our work follows this line of expressive sound synthesis but we focus on acoustic guitar. In particular, we propose guitarroll, a new input representation for guitar synthesis models. This is inspired by the tablature score that provides the finger position of the played notes. Unlike the pianoroll representation which is highly sparse, guitarroll provides more efficient and suitable input representation for guitar synthesis models.

\def\hide{our proposed system receives MIDI rather than musical score for acoustic guitar. Also, our work focuses on generating expressive audio which we define as the discrepancy between audio synthesized by sample-based synthesis and real recordings.}

We employ a diffusion-based model to synthesize expressive guitar sounds conditioned on the guitarroll representation. Diffusion-based generative models have enabled neural networks to model complex data distributions with high stability \cite{ho2020ddpm}. Recent work has demonstrated their capability to synthesize high-quality instrumental sounds, including a mixture of multiple instrument tracks \cite{Hawthorne2022multiinstdiff}. A noticeable feature of this work is that the model concatenates the feature of the previously generated output along the conditioning input MIDI feature to achieve coherence along the long-term audio waveform. However, directly conditioning a feature consisting of different modalities and temporal information is inefficient for the model. 

Our proposed model is based on \cite{Hawthorne2022multiinstdiff}, but we streamline the model in two ways. First, we implement a diffusion outpainting algorithm to generate coherent long-term audio waveform, eliminating the need of inefficient concatenation. Second, we downsize the model to align with our single-instrument synthesis task. Given the limited availability of real-world performance data, the GuitarSet dataset \cite{xi2018guitarset}, comprising only 3 hours of excerpts, prompted us to construct an additional, large-scale MIDI/audio-aligned acoustic guitar dataset for pre-training, utilizing off-the-shelf virtual instruments. The outpainting algorithm has led to improvement in both objective and subjective metrics compared to the baseline approach and remains comparable with the previous large-scale model \cite{Hawthorne2022multiinstdiff}, which concurrently trained instruments using both real-world and sample-based synthesized audio. Audio examples are available online\footnote{https://hanshounsu.github.io/guitar-synthesis-diffusion-outpainting/}.

\def\hide{This paper's contributions are twofold: 1) We suggest a novel and efficient representation method which we call guitarroll. 2) We implement the diffusion outpainting algorithm to generate long-term audio coherently, which we claim to be more efficient than the architecture suggested in \cite{Hawthorne2022multiinstdiff}. Audio examples are provided online \footnote{https://hanshounsu.github.io/guitar-synthesis-diffusion-outpainting/}.}

\section{Method}
\label{sec:method}

\subsection{Overall Architecture}
\label{ssec:method_1}

The overall architecture is based on the T5 structure \cite{raffel2020t5}, where the denoising target is the mel spectrogram. In addition, Soundstream vocoder \cite{zeghidour2021soundstream} is responsible for translating mel spectrograms into audio waveforms. The T5 decoder performs the denoising task with the encoder providing the MIDI condition. The overall model size is downscaled, and an additional encoder providing features of the previous mel spectrogram is removed, which leads to a simpler structure. We do not discuss the vocoder in detail, as we used the same pre-trained model distributed by \cite{Hawthorne2022multiinstdiff}\footnote{https://tfhub.dev/google/soundstream/mel/decoder/music/1}.

\subsection{The Guitarroll Representation for Guitars}
\label{ssec:method_2}

\begin{figure}[!]
\begin{minipage}[b]{1.0\linewidth}
  \centering
  \centerline{\includegraphics[width=7.5cm]{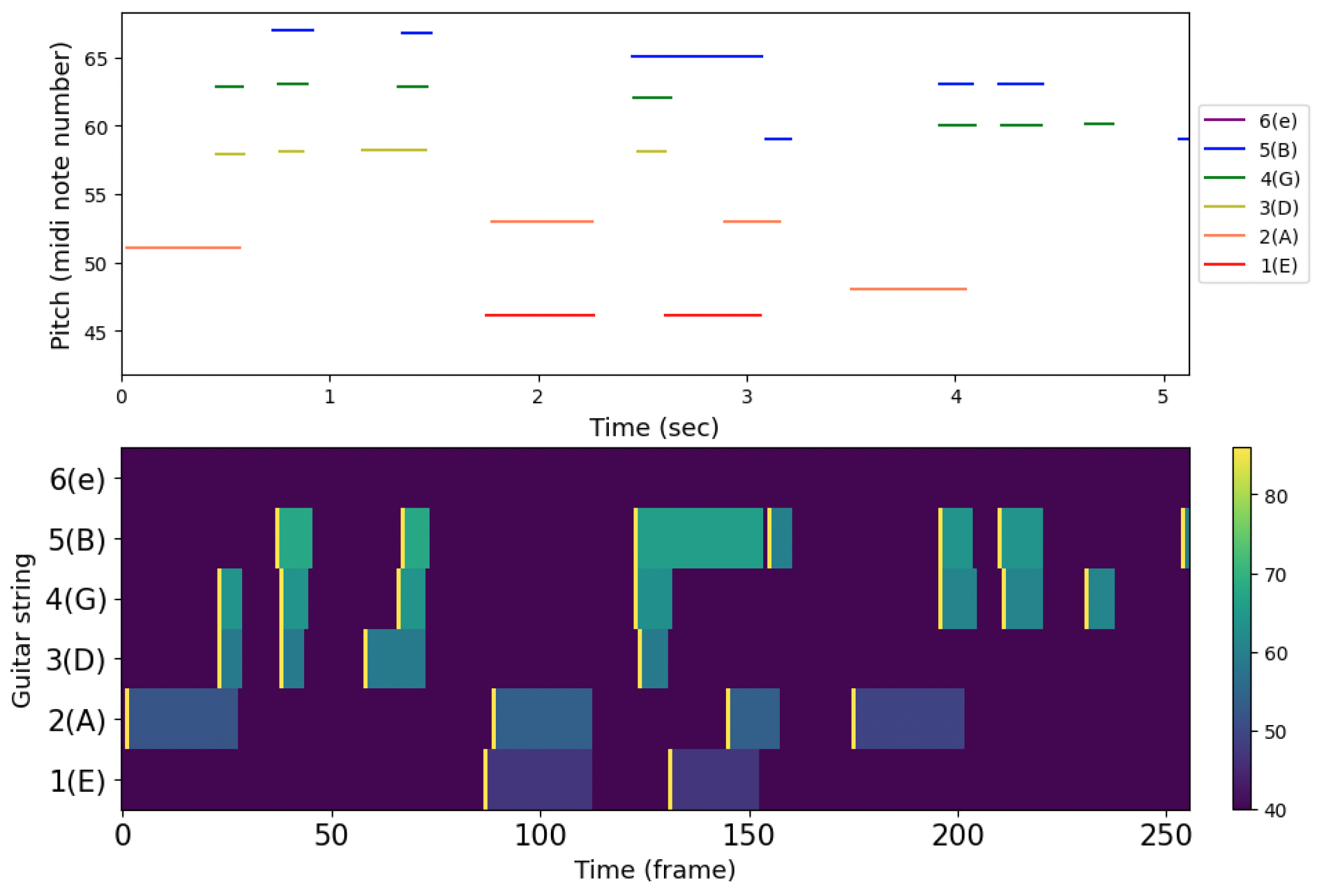}}
\end{minipage}
\caption{An MIDI plot of GuitarSet \cite{xi2018guitarset} (upper) and the corresponding guitarroll representation (bottom). The note pitch ranges from 40 (E1, dark purple) to 84 (C6, bright yellow). The number corresponding to the onset position (85) is noticeable as yellow vertical lines in each note's starting point.} 
\label{fig:jams_midi_rep}
\end{figure}

Previous works represent MIDI as a pianoroll \cite{wang2019performancenet, kimjongwook2019, dong2022deepperformer}, or MIDI-like tokens \cite{Hawthorne2022multiinstdiff}. However, the pianoroll is sparse as most of the note pitch are not usually activated. This leads unnecessary high dimensionality. The MIDI-like tokens imposes excessive embeddings to learn such as those for the time events. In our approach, we propose \textit{guitarroll}, a matrix consisting of size-6 vectors, each containing the note pitch number for each time frame, as depicted in Figure \ref{fig:jams_midi_rep}. The note pitch numbers are embedded into a higher dimension of $D$ when input to the model, eventually resulting in a $6D \times L$ matrix. This approach allows us to furnish the model with a denser representation while explicitly providing temporal information to the model.

\begin{figure}[t]
\begin{minipage}[b]{1.0\linewidth}
  \centering
  \centerline{\includegraphics[width=6.5cm]{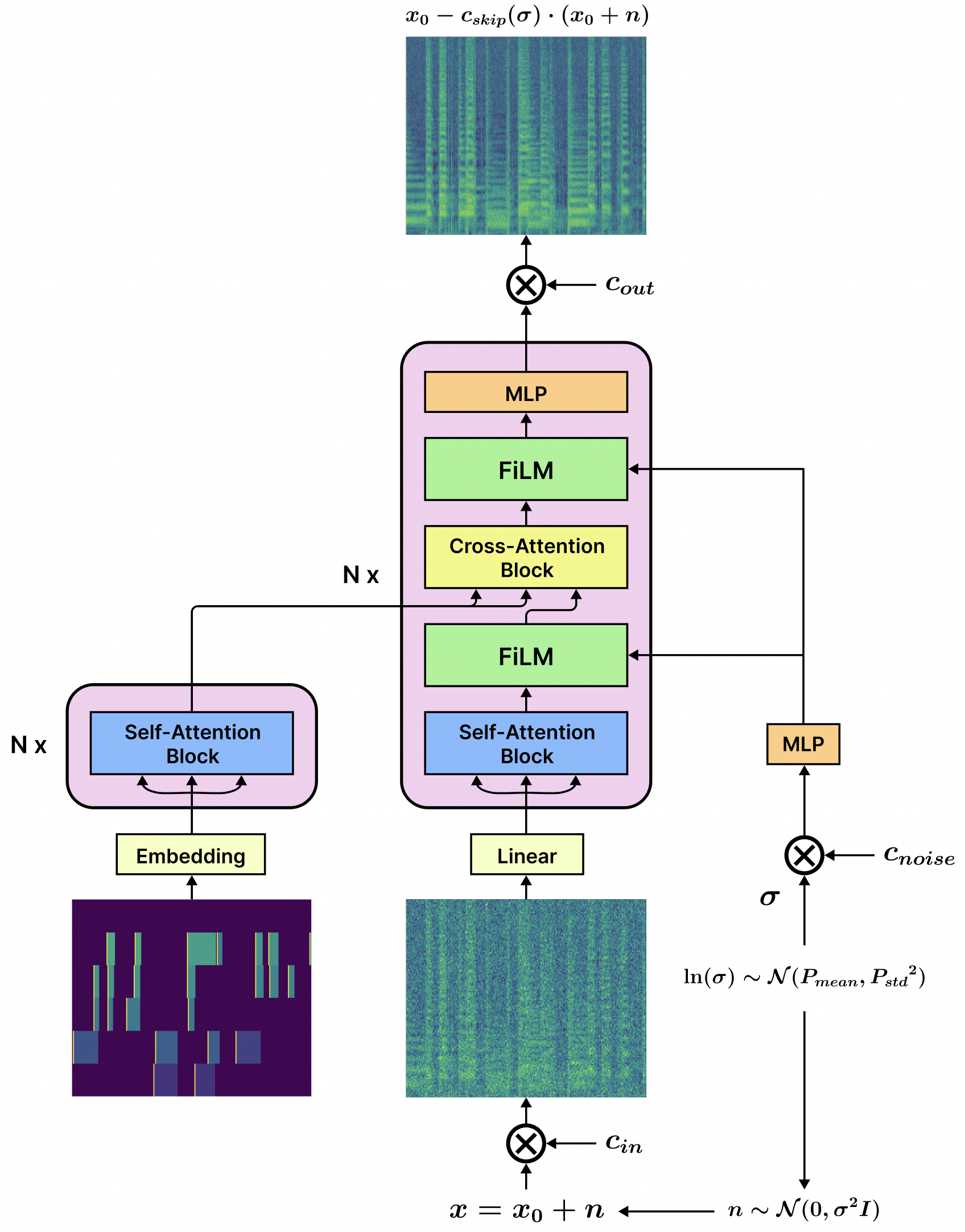}}
\end{minipage}
\caption{Overall model architecture. $c_{in}, c_{out}, c_{noise}, c_{skip}$ are the scaling schedules, which are dependent on $\sigma_{data}=0.1$. $\rho=9, P_{mean}=-3.0, P_{std}=1.0$. Details on the definition of each variable are described at \cite{Karras2022edm}.}
\label{fig:overall_architecture}
\end{figure}

\subsection{Diffusion-based Generative Modeling}
\label{ssec:method_3}

Diffusion models consist of a forward diffusion process where a small amount of Gaussian noise is gradually applied to the data and a reverse process where the true sample is recreated using a denoising neural network. The final objective of our diffusion model is learning the score function $\nabla_{\bm{M_t}}\text{log}q_t(\bm{M}_t| \bm{c})$ in order to sample data distribution $p_{\mathrm{data}}(\mathbf{M_0},\mathbf{c})$, where $\bm{c}$ is the MIDI condition and $\bm{M}_0$ denotes the corresponding mel spectrogram.

Due to the transition property of the forward process, the distribution of perturbed data $\bm{M}_t$ is described as follows:
\begin{equation}
q_{0t}(\bm{M}_t|\bm{M}_0)=\mathcal{N}(\bm{M}_t|\alpha_t\bm{M}_0,\sigma_t^2\mathbf{I})
\end{equation}
\noindent where $\sigma_t$ and $\alpha_t$ are noise schedules,
and time step $t \in [0, T]$ is set as a monotonic function $f$ of $\sigma_t$, i.e., $t=f^{-1}(\sigma_t)$ where $T>0$.

Previous leading works on diffusion-based model \cite{ho2020ddpm, song2021scorebased, Karras2022edm} note a strong connection between the training objectives of diffusion models and score-based generative models. Their discrepancy is minimized when the diffusion step is infinitesimal. In this case, the forward process can be expressed as a stochastic differential equation (SDE) \cite{song2021scorebased}:
\begin{equation}
d\bm{M}_t = f(t)\bm{M}_tdt + g(t)d\bm{w}_t, \;\;\; \bm{M}_0 \sim q_0(\bm{M}_0)
\end{equation}
where $w_t \in \mathbb{R}^D$ is the standard Wiener process, and $f$ and $g$ are the drift and diffusion coefficient respectively.
The reverse process can be expressed as a reverse-time SDE:
\begin{equation}
d\bm{M}_t = [f(t)\bm{M}_t-g^2(t)\nabla_{\bm{M}_t}\text{log}q_t(\bm{M}_t| \bm{c})]dt + g(t)d\bm{\bar{w}}_t,
\end{equation}
where $\nabla_{\bm{M}_t}\text{log}q_t(\bm{M}_t| \bm{c})]dt$ is the score function, $\bm{M}_T \sim q_T(\bm{M}_T)$ and $\bar{w_t} \in \mathbb{R}^D$ is the standard Wiener process in reverse time.

The objective can be expressed as follows:
\begin{equation}
\begin{split}
\mathcal{L}(\theta; \omega(t)) & = \frac{1}{2}\int^{T}_{0} \omega(t)\mathbb{E}_{q_t(\bm{M}_t)}[||\bm{\epsilon}_\theta(\bm{M}_t,\bm{c},t) \\
& \qquad\qquad\qquad\qquad+\sigma_t\nabla_{\bm{M}_t}\mathrm{log}q_t(\bm{M}_t|\bm{c})||^2_2] \mathrm{d}t \\
 & = \frac{1}{2}\int^{T}_{0} \omega(t)\mathbb{E}_{q_0(\bm{M}_0)}\mathbb{E}_{q(\bm{\epsilon})}[||\bm{\epsilon}_\theta(\bm{M}_t,\bm{c},t) \\
 & \qquad\qquad\qquad\qquad\qquad -\bm{\epsilon}||^2_2]\mathrm{d}t+C, 
\end{split}
\end{equation}
where $\omega(t)$ is the weight term, and $\bm{\epsilon}_\theta(\bm{x},t)$ is the model prediction. In our work, we adopt the training and sampling method of recent work \cite{Karras2022edm} for our training scheme, as illustrated in Figure \ref{fig:overall_architecture} and Algorithm \ref{outpainting algorithm}. In the reverse diffusion sampling process, we generate samples with 40 steps by applying the $2^{\text{nd}}$ order Runge-Kutta method on the ancestral sampling algorithm, roughly following the publicly available repository\footnote{https://github.com/archinetai/audio-diffusion-pytorch/tree/v0.0.90}.

\subsection{Diffusion Outpainting Method For Continuation}
\label{ssec:method_3}
The parallel sampling procedure inherent in the diffusion decoder, as described in \cite{Hawthorne2022multiinstdiff}, poses high computational demands, particularly when dealing with long sequences.  They attempted to overcome this issue by generating output in a segment-by-segment manner, with the previous output segment conditioned through a context encoder. However, this issue can be more efficiently resolved by \textit{outpainting}, where the model generates open-ended content of cropped data.

There have been active works regarding the diffusion-based approach to solving linear inverse problems, and we employ unsupervised approach which fulfills inverse tasks without additional conditioning modules. Current works such as \cite{chung2022manifold, lugmayer2022repaint, kawar2022ddrm} denote the robustness of their unsupervised approach compared to existing state-of-the-art approaches. Among these works, we adopt \textit{RePaint} \cite{lugmayer2022repaint} due to its simplicity and robust performance. The detailed outpainting procedure is depicted in Algorithm \ref{outpainting algorithm}.

\begin{algorithm}[t]
\setstretch{1.15}
    \caption{Outpainting Algorithm.} \label{outpainting algorithm}
  \begin{algorithmic}[1]
    \REQUIRE $\{t_i\}^N_{i=1}$, $\bm{\epsilon}_\theta$, $\bm{c}$, $m = [\smash[b]{\underbrace{\mathbbm{1}_n \dots \mathbbm{1}_n}_{\text{\tiny L/2 times }}}, \smash[b]{\underbrace{\vec{\mathbf{0}}_n \dots \vec{\mathbf{0}}_n}_{\text{\tiny L/2 times}}}]$ 
    \STATE $\hat{\bm{M}}_N \sim \mathcal{N}(\bm{0},\bm{I})$
    \FOR {$i = N , ..., 1$}
      \FOR {$u = 1 , ..., U$}
        \STATE $\bm{\epsilon} \sim \mathcal{N}(\mathbf{0}, \mathbf{I})$ if $i > 1$, else $\bm{\epsilon}=0$,
        \STATE $\bm{M}_{i-1}^{0:0.5L} = m \odot (\bm{M}_0 + \sigma_{t_{i-1}}\bm{\epsilon})$
        \STATE $\hat{\bm{M}}_{i-1} \xleftarrow[\texttt{Sampling Step}]{\texttt{RK2}} \hat{\bm{M}}_i, \bm{\epsilon}_\theta, \bm{c}, \sigma_{t_i}, \sigma_{t_{i-1}}, \bm{z} \sim \mathcal{N}(\mathbf{0}, \mathbf{I})$
        \STATE $\hat{\bm{M}}_{i-1}^{0.5L:L} = (1-m) \odot \hat{\bm{M}}_{i-1}$
        \STATE $\hat{\bm{M}}_{i-1} = \bm{M}_{i-1}^{0:0.5L} + \hat{\bm{M}}_{i-1}^{0.5L:L}$
        \IF {$u<U$ and $i>1$}
          \STATE $\beta_{i-1} = \sqrt{\sigma_{t_i}^2 - \sigma_{t_{i-1}}^2}$
          \STATE $\hat{\bm{M}}_i \sim \mathcal{N}(\hat{\bm{M}}_{i-1}, \beta_{i-1}\bm{I})$
        \ENDIF
      \ENDFOR
    \ENDFOR
    \STATE \textbf{return} {$\hat{\bm{M}}_0$}
  \end{algorithmic}
\end{algorithm}

\begin{figure}[t]
\begin{minipage}[b]{1.0\linewidth}
  \centering
  \centerline{\includegraphics[width=7cm]{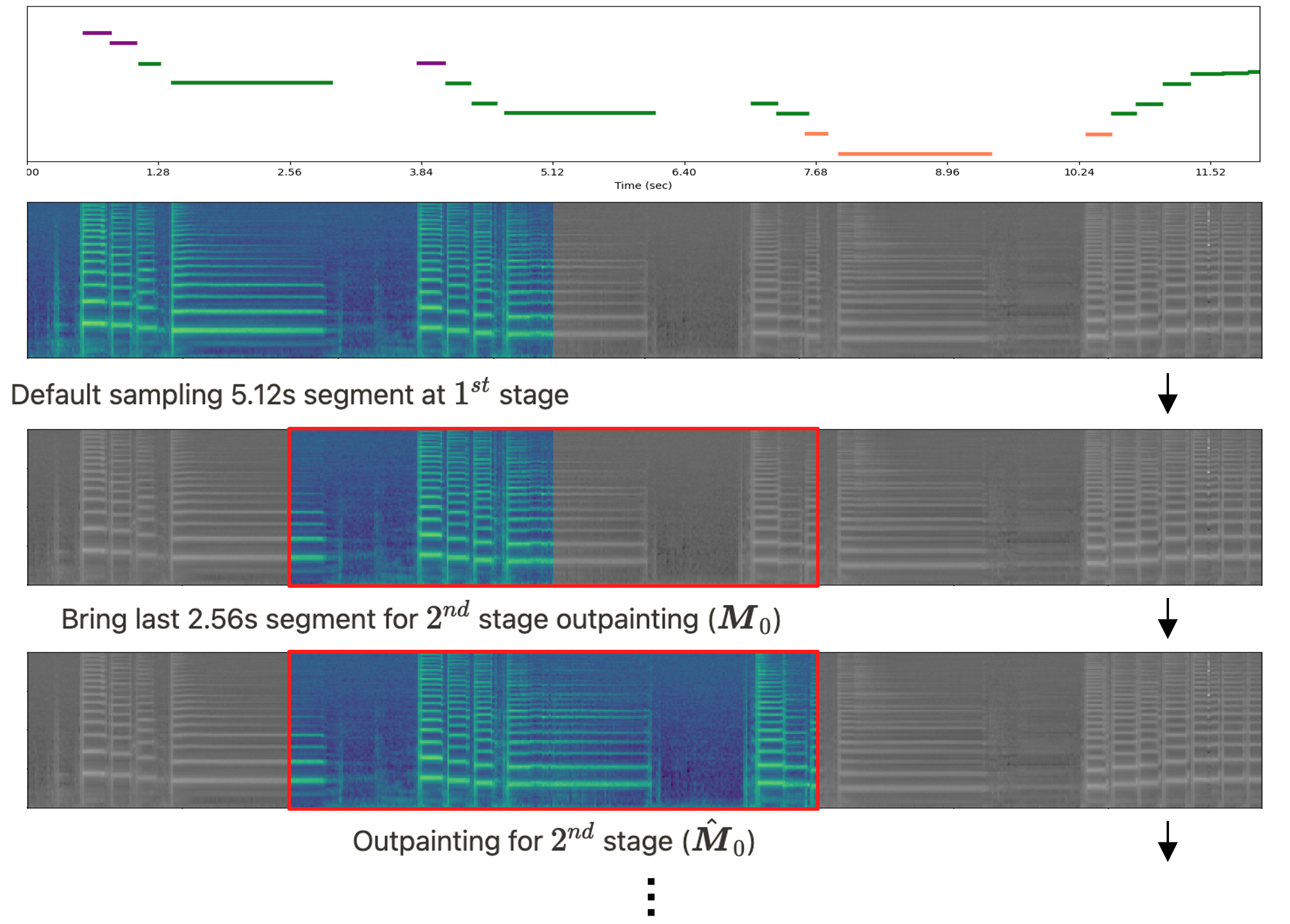}}
  \vspace{-0.0cm}
\end{minipage}
\vspace{-0.1cm}
\caption{ The proposed outpainting algorithm for the acoustic guitar synthesis task. In order to generate coherent audio results from a long MIDI note sequence, starting from the $2^{\text{nd}}$ stage the model attempts to fill in the remainder when given previously generated data.}
\vspace{0.0cm}
\label{fig:inpainting_algorithm}
\end{figure}

\begin{table*}[htbp]
  \renewcommand{\arraystretch}{1}
  \centering
  \begin{tabular}{c|c|c c c|c c c|c c c}
    \Xhline{2\arrayrulewidth}
    \multirow{2}{*}{Model} & \multirow{2}{*}{Parameters} & \multicolumn{3}{c|}{VGGish FAD $\downarrow$} & \multicolumn{3}{c|}{PANNs FAD $\downarrow$} & \multicolumn{3}{c}{Transcription F1 $\uparrow$}\\
    & & all & comp & solo & all & comp & solo &all & comp & solo \\
    \hline
    BaselineFT & 41.2M & 4.72 & 4.85 & 4.92 & 35.49 & 32.68 & 42.08 & 0.484 & 0.330 & 0.638\\
    OutpaintFT-pianoroll & 35.5M & 5.82 & 5.40 & 6.62 & 49.85 & 39.43 & 66.44 & - & - & - \\
    OutpaintFT-guitarroll (ours) & 35.5M & \textbf{4.27} & \textbf{4.16} & \textbf{4.66} & \textbf{33.82} & \textbf{29.53} & \textbf{41.93} & \textbf{0.531} & 0.375 & \textbf{0.688} \\
    Hawthorne Demo \cite{Hawthorne2022multiinstdiff} & 412M & 4.72 & 4.91 & 5.03 & 41.44 & 35.02 & 52.86 & 0.530 & \textbf{0.470} & 0.590\\
    \hline
    Ground Truth Encoded & - & 3.03 & 2.74 & 3.53 & 21.69 & 16.05 & 30.61 & 0.569 & 0.461 & 0.678 \\
    Ground Truth Raw & - & 0.088 & 0.096 & 0.22 & 3.30 & 4.20 & 4.52 & 0.781 & 0.733 & 0.828  \\
    \Xhline{2\arrayrulewidth}
  \end{tabular}
  \caption[Results on objective metrics]{Objective evaluation results on FAD and Transcription F1 score. The metrics are averaged along the evaluation set. Ground Truth Encoded denotes the output audio when the ground-truth mel spectrogram is decoded through the Soundstream pre-trained vocoder, and Ground Truth Raw denotes the ground-truth audio.}
  \label{objective}
\end{table*}

\begin{table*}[t]
  \renewcommand{\arraystretch}{1.1}
  \centering
  \setlength{\tabcolsep}{3.5pt}
  \begin{tabular}{c|cc|cc|cc|cc}
    \Xhline{2\arrayrulewidth}
    \multirow{2}{*}{Model} & \multicolumn{2}{c|}{Timbre} & \multicolumn{2}{c|}{Note accuracy} & \multicolumn{2}{c|}{Expressiveness} & \multicolumn{2}{c}{Overall Quality} \\
    & comp & solo & comp & solo & comp & solo & comp & solo \\
    \hline
    BaselineFT & 2.66$\pm$.07 & 2.82$\pm$.07 & 3.12$\pm$.09 & 3.34$\pm$.08 & 2.86$\pm$.08 & 2.88$\pm$.08 & 2.76$\pm$.06 & 2.89$\pm$.06 \\
    OutpaintFT (ours) & \textbf{2.97}$\pm$.07 & \textbf{3.10}$\pm$.06 & 3.46$\pm$.07 & 3.61$\pm$.08 & 3.12$\pm$.08 & 3.17$\pm$.07 & 3.05$\pm$.06 & \textbf{3.19}$\pm$.06 \\
    Hawthorne Demo \cite{Hawthorne2022multiinstdiff} & 2.85$\pm$.07 & 2.80$\pm$.07 & \textbf{3.64}$\pm$.07 & 3.61$\pm$.08 & \textbf{3.30}$\pm$.07 & \textbf{3.26}$\pm$.08 & \textbf{3.16}$\pm$.06 & 3.10$\pm$.06 \\
    \hline
    Ground Truth Raw & 4.81$\pm$.04 & 4.84$\pm$.03 & 4.81$\pm$.04 & 4.83$\pm$.03 & 4.75$\pm$.04 & 4.77$\pm$.03 & 4.77$\pm$.04 & 4.81$\pm$.03 \\
    \hline
    \Xhline{2\arrayrulewidth}
  \end{tabular}
  \caption[5-point Likert scale MOS results on the baseline and proposed model.]{5-point Likert scale MOS results on the baseline and proposed model.}
  \label{MOS}
\end{table*}

\section{Experiments}
\label{sec:experiments}

\subsection{Datasets}
\label{sec:experiments_1}

Our model is trained using the GuitarSet dataset \cite{xi2018guitarset}, which comprises 3 hours of realistic acoustic guitar excerpts played by six experienced guitarists, accompanied by corresponding MIDI annotations. The excerpts consist of \textit{comp} samples and \textit{solo} samples, where each \textit{comp} sample represents a chord accompaniment for a corresponding \textit{solo} sample. The dataset contains onset, offset, note pitch and string number, but no velocity information. We use the same train/test split as \cite{Hawthorne2022multiinstdiff}, where the third progressions of each style are used as test tracks, which leads to 2 hours for training and 1 hour of test.

Initially, we attempted to train solely on GuitarSet, but its limited quantity yielded unsatisfactory outputs. Therefore, we constructed a large MIDI/audio-aligned acoustic guitar dataset for pre-training, obtained by sample-based method. This dataset, which we refer to as the \textit{Lakh-Ilya dataset}, draws MIDI data from the Lakh MIDI Dataset \cite{Raffel2016LearningBasedMF} and aligns it with audio synthesized by an acoustic guitar library provided by an established production company\footnote{https://www.ilyaefimov.com/products/acoustic-guitars/acoustic-guitar.html}.  The dataset consists of more than 100 hours of data and was split by allocating 90\% of tracks for training and 10\% of tracks for validation.

\subsection{Experiment Settings}
\label{sec:experiments_2}

We downsample the audio to 16kHz and adopt an FFT size of 1024 samples, a hop size of 320 samples, a frame size of 640 samples and a mel bin size of 128 to obtain the mel spectrogram. This leads to a total of 256 frames for the 5.12s input segment. The MIDI representation is frame-wise aligned to the mel spectrogram. We fix the model specification as follows: 6 FFT layers for each decoder and encoder block, 256 channel size for each block and embeddings, 3 attention heads with 128 channel dimensions each, 1024 channel size for FFNN blocks, and 512 channel size for MLP blocks. This leads to a total 35.5M trainable parameters. We applied Classifier-Free Guidance \cite{ho2021classifierfree} for conditioning the MIDI representation. The model is initially pre-trained with the Lakh-Ilya dataset for 400K steps, then fine-tuned with GuitarSet for 300K steps. We apply a learning rate of 1e-4, a gradient accumulation of 2, a batch size of 64, a dropout of 0.1, and the Adam optimizer for the two training phases. Training took around 2$\sim$3 days for each phase. The outpainting ratio was set to 0.5, where we generated 2.56s of mel spectrogram when given 2.56s of the previously generated audio.

\subsection{Baselines}

To validate the efficacy of our proposed architecture (OutpaintFT), we compared it to the architecture based on \cite{Hawthorne2022multiinstdiff}, with the model specification downsized to match our model, denoted as BaselineFT. We also attempted to verify the effectiveness of our proposed MIDI representation (guitarroll) compared to the previous approach (pianoroll). However, the output quality was subpar, with no clear melody aligned with the input MIDI. Therefore we only display partial, obtainable results with the pianoroll representation, as shown in Table \ref{objective}. Though it is not a fair comparison, we also displayed synthesized results of \cite{Hawthorne2022multiinstdiff} (Hawthorne Demo) obtained from the demo website\footnote{https://github.com/magenta/music-spectrogram-diffusion}.

\subsection{Objective Metrics}

For the objective evaluations, we included the Fr$\acute{\text{e}}$chet Audio Distance (FAD) \cite{kilgour2019fad} from VGGish \cite{hershey2017vggish} and PANNs \cite{kong2020panns} models, and F1 score from MT3 \cite{gardner2022mt3} model. FAD assesses the perceptual similarity between the ground-truth set and the generated set, specifically focusing on timbral quality in our case. MT3, a multi-track transcription model, predicts transcribed MIDI from audio containing multiple instruments. By passing the audio generated by our trained model through a pre-trained MT3 transcription model, we obtained transcribed MIDI. The similarity between the original MIDI and the transcribed MIDI indicates the degree to which our model accurately follows the input MIDI notes. Due to our task being single instrument synthesis, we report F1 scores on the ``Flat'' group level, which disregards instrument program number prediction. The results are also analyzed on solo and comp sets to measure performance of the model on both simple monophonic samples and complex polyphonic samples.

\subsection{Listening Test}

We conducted a crowdsourced listening test with \textcolor{black}{17} participants over 8 samples from the comp set and 8 samples from the solo set. 6-second samples are randomly cropped from the test set. Participants were instructed to rate the audio output on a 5-point Likert scale, with metrics instructed as follows:

\begin{itemize}
    \item \textbf{Timbre:} Quality of the output sound's timbre.
    \item \textbf{Note accuracy:} Whether the note pitch is detuned or the note sound is not clear.
    \item \textbf{Expressiveness:} Whether the output sounds close to a realistic acoustic guitar performance rather than a virtual instrument.
    \item \textbf{Overall quality:} Overall quality of the output sound.
\end{itemize}

We evaluated timbre, note accuracy, and expressiveness metrics based on their relative similarity to ground-truth results, while the overall quality metric was assessed through absolute evaluation.

\section{Results}
\label{sec:results}

\subsection{Quantitative Results}
\label{ssec:results_1}

The objective evaluation results are denoted in Table \ref{objective}. Our suggested model architecture (OutpaintFT-guitarroll) showed better results on FAD measures, outperforming the previous work (Hawthorne Demo), baseline model (BaselineFT), and OutpaintFT-pianoroll model. In the case of Transcription F1 score, our proposed model achieved higher results across all sets compared to the baseline model, but compared to the previous work, results on the comp set were relatively poor.  This discrepancy can be attributed to the fact that the previous work was also trained on the Slakh2100 dataset \cite{manilow2019slakh}, which contains sample-based synthesized audio, and thus contains finer envelopes.

\subsection{Qualitative Results}
\label{ssec:results_2}

The mean and standard error values for the listening test are summarized in Table \ref{MOS}. OutpaintFT performs better than BaselineFT across all evaluation metrics and is generally comparable to the results from Hawthorne Demo. Notably, the timbre metric was higher than Hawthorne Demo as expected. However, the expressive metric was relatively poor, along with the note accuracy on the comp set, due to the same reason described in \ref{ssec:results_1}.
 
\section{Conclusion}
\label{sec:conclusion}

We presented a diffusion-based MIDI-to-audio synthesis network for synthesizing expressive acoustic guitar sound. We pointed out the inefficiency of the previous architecture and proposed a simpler architecture that generates audio segments using diffusion outpainting. Our model also utilizes an efficient input representation suitable for acoustic guitar instruments. Objective and subjective evaluation results indicate that our architecture outperforms the baseline structure and synthesizes more realistic timbre than the previous work. While these findings are promising, we acknowledge that further enhancements in overall quality could be explored in the future.

\vfill\pagebreak

\bibliographystyle{IEEEbib}
\bibliography{strings,refs}

\end{document}